\documentclass[iop,apj]{emulateapj}
\usepackage{graphicx}
\usepackage{amssymb}
\usepackage{amsmath}
\usepackage{mathrsfs}
\usepackage{subeqnarray}
\shorttitle{physical origin of the variability related correlation}

\begin{document}
\title{What can we learn about GRB from the variability timescale related correlations?}
\author{Wei Xie$^{1}$,  Wei-Hua Lei*$^{1}$, Ding-Xiong Wang$^{1}$}
\affil{$^{1}$School of Physics, Huazhong University of Science and Technology, Wuhan 430074, China. Email: leiwh@hust.edu.cn
}

\begin{abstract}
Recently, two empirical correlations related to the minimum variability timescale ($\rm MTS$) of the lightcures are discovered in gamma-ray bursts (GRBs). One is the anti-correlation between $\rm MTS$ and Lorentz factor $\Gamma$, the other is the anti-correlation between the $\rm MTS$ and gamma-ray luminosity $L_\gamma$. Both the two correlations might be used to explore the activity of the central engine of GRBs. In this paper we try to understand these empirical correlations by combining two popular black hole (BH) central engine models (namely, Blandford \& Znajek mechanism and neutrino-dominated accretion flow). By taking the $\rm MTS$ as the timescale of viscous instability of the neutrino-dominated accretion flow (NDAF), we find that these correlations favor the scenario in which the jet is driven by Blandford-Znajek (BZ) mechanism.
\end{abstract}

\section{Introduction}

The mechanism for launching a relativistic jet from gamma-ray burst (GRB) central engine is still unclear. The leading GRB central engine model involves a stellar mass black hole (BH) surrounded by a hyper-accreting disk. An alternative model involves a rapidly spinning, strongly magnetized neutron star (also known as ``millisecond magnetar", e.g., \citealt{Usov92, LZ14}). In the BH scenario, there are two main energy reservoirs to provide the jet power: the gravitational energy in the neutrino dominated accretion flow (NDAF) that is carried by neutrinos and anti-neutrinos, which annihilate and power a bipolar outflow ( \citealt{PWF99, NPK01, DPN02, KM02, Gu06, Liu07, Liu15, CB07, JYPD07, Lei08, Lei09, Lei13, Yi17}); and the spin energy of the BH which can be tapped by a magnetic field connecting a remote astrophysical load through the Blandford \& Znajek mechanism (\citealt[hereafter BZ]{BZ77}). It's hard to directly distinguish these different models, since no radiation (except gravitational wave and neutrinos) reaches the observer directly from the central engine. However, the empirical correlations of some observational variables (e.g. the Lorentz-factor--isotropic-luminosity/energy correlations (\citealt{Liang10,Gld12,LZ12}), the Lorentz-factor--beaming-corrected-energy/luminosity correlations(\citealt{Yi16})), may place constrains on GRB central engine models.

The prompt emission of GRBs are extremely variable. There are two empirical correlations involed with the minimum variability timescale ($\rm MTS$, a measurement of the temporal variability of the lightcures) proposed by \cite{Sonbas15}. One is the anti-correlation between ${\rm MTS}$ and the bulk Lorentz factor $\Gamma$. The other is the anti-correlation between ${\rm MTS}$ and the isotropic gamma-ray luminosity $L_\gamma$. With adopting the MTS measurements determined by the structure-function method (\citealt{GB14}), and the Lorentz factors estimated from the conventional method which takes the peak of the early afterglow light curve as the deceleration time of the external foward shock (\citealt{LZ12, Liang15}), \cite{Wu16a} confirmed these correlations and extended them to the blazars, with results as ${\rm MTS} \propto \Gamma^{- 4.8 \pm 1.5}$ and $\mathrm{MTS}\propto L_\gamma^{-1.0\pm0.1}$. All these parameters (${\rm MTS}$, $\Gamma$ and $L_\gamma$) are closely related to the central engine, the correlations here are therefore expected to shed light on the physics about jet acceleration and the accreting activity.

The origin of the MTS is on debate. One scenario proposes that the variation time scale hints the dissipation process related to the turbulence or magnetic reconnection in the jet itself (\citealt{NK09}). The other possibility is that the variation timescale reflects the time intervals of the internal shocks modulated by the central engine activity (\citealt{KPS97}). A possible process causing the tempral variation is the instability ocurred in the disk. The thermal instability, for example, appearing in the inner region of the geometrically thin and optically thick accretion disk, has been used to explain the `heartbeat'-like oscillations observed in some XRBs and AGNs (\citealt{GKS15, Wu16b}). NDAF has long been thought to be stable under parameters of interest (\citealt{DPN02}). However, based on the detailed treatment of the chemical equilibrium in the gas species, \cite{JYPD07} argued that NDAF can be viscously and thermally unstable at extremely high accretion rate. Moreover, \cite{Lei09} proposed that the NDAF could be viscouslly unstable at more moderate accretion rate when considering certain magnetic mechanism such as the magnetic coupling (MC) between the plumping region and the disk. More recently, \cite{XLW16} argued that a none-zero inner boundary torque should be considered. This revised NDAF was found to be viscously unstable.

In this paper, we attribute the temporal variability $\rm MTS$ of the GRBs to the viscous instability of the NDAF. Meanwhile, on the basis of the investigation of the Lorentz factor of the GRB jet separatively driven by BZ and NDAF models, we can infer that which jet production process is possible by checking whether it can reproduce these two $\rm MTS$-related correlations.

This paper is organized as follows. In Section 2, we derive the $\rm MTS$ from the viscous timescale of the unstable NDAF, and then calculate the Lorentz factor of the jet driven by NDAF and BZ mechanisms, respectively. The theoretical predictions for the ${\rm MTS}-\Gamma$ as well as ${\rm MTS}-L_\gamma$ relation are compared with the empirical results. In Section 3, we summarize our conclusion and discuss the implications of the empirical $\rm MTS$ related correlations in the GRB central engine.

\section{The BH Central Engine Model and MTS-$\Gamma$ correlations}
A hyper-accretion stellar BH system is the prevailing GRB central engine model. We summarize the main features of this model, focusing on the timescale of the viscous instability in NDAF, the baryon loading and power of the jet. We equate $\rm MTS$ to the timescale of the instability. The initial Lorentz factors and total power of the jets in NDAF and BZ mechanisms are calculated in the same way as in \cite{Lei13}.

\begin{figure}[h!t!]
\centering
\includegraphics[angle=0,scale=0.4]{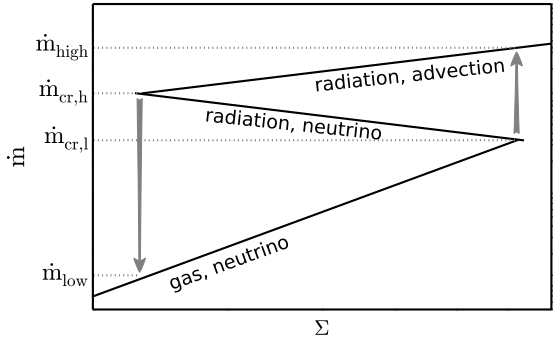}
\caption{The $\dot{m}-\Sigma$ profile of the unstable region in the inner side of the NDAF, the axes are scale-free. The two arrows denote the state transition between the ``high state" and ``low state" of the unstable region.}
\label{fig:S-curve}
\end{figure}

\subsection{MTS and the viscous instability of NDAF}

According to the previous works (\citealt{XLW16, Lei09}), the viscous instability can occur in the inner region of the NDAF when considering the plausible extra magnetic torque. Furthermore, the whole disk can be divided into five regions successively from the outer side to the inner side: region I (radiation-pressure dominated ADAF), region II (gas-pressure dominated, transparent NDAF), region III (gas-pressure dominated, opaque NDAF), region IV (radiation-pressure dominated, opaque NDAF) and region V (radiation-pressure dominated ADAF).

Firstly, we'd like to estimate the location of the unstable region from the fact that the unstable region (region IV) is distinguished from its upper stream (region III) when the radiation pressure becomes import. The fraction of the radiation pressure in region III is expressed as
\begin{equation}
P_\mathrm{rad}/P=6.6\times10^{2}A^{-1}B^{-3/4}C^{-1/4}\mathscr{D}^{2} \alpha^{-1/2}m^{-5/2}\dot{m}^{2}R^{-15/4},
\label{eq:pradp}
\end{equation}
where $P$ is the total pressure, $m\equiv M_\bullet/M_\odot$ is the normalized BH mass, and $\dot m\equiv \dot{M}/M_\odot \rm s^{-1}$ is the disk accretion rate in unit of $M_\odot \rm s^{-1}$. $A$, $B$, $C$ and $D$ are the relativistic correction factors for a disk around a Kerr BH \citep{RH95}, i.e.,
\begin{subequations}
\begin{align}
& A = 1 - 2(r/r_{\rm g})^{-1} + a_\bullet^2 (r/r_{\rm g})^{-2}, \\
& B = 1 - 3(r/r_{\rm g})^{-1} + 2a_\bullet (r/r_{\rm g})^{-3 / 2}, \\
& C = 1 - 4a_\bullet (r/r_{\rm g})^{-3 / 2} + 3 a_\bullet^2 (r/r_{\rm g})^{-2}, \\
& D = \int_{r_{\rm{ms}}/r_{\rm g}}^{r/r_{\rm g}} {\frac{ x^2 - 6 x + 8 a_\bullet x^{1/2} - 3 a_\bullet ^2 }{  2 \sqrt{Rx} (x^2 - 3x + 2a_\bullet x^{1/2} )}dx},
\end{align}
\label{KerrFactors}
\end{subequations}
where $r_{\rm g} \equiv G M_\bullet /c^2$ denotes the gravitational radius, and $r_{\rm ms}$ is the radius of the innermost stable circular orbit (ISCO). The symbol $\mathscr{D}$ in equation (\ref{eq:pradp}) is defined as $\mathscr{D}\equiv D+\eta A_{\mathrm ms}L_{\mathrm ms}/r^2\Omega_\mathrm{k}$, in which $\eta$ is introduced to parameterize the magnitude of the extra magnetic torque, $L_{\rm{ms}}=2GM(3\chi_\text{ms}-2a_\bullet)/\sqrt{3}c\chi_\text{ms}$ is the specific angular momentum of a particle in the disk, and $\chi_\text{ms}\equiv\sqrt{c^2r_\text{ms}/GM}$. According to \cite{XLW16}, NDAF becomes unstable at the inner region when a significant non-zero boundary torque is applied. In the unstable region, the coefficients in equation (\ref{eq:pradp}) can be approximately as $A^{-1}B^{-3/4}C^{-1/4}\mathscr{D}^{2}\simeq8.5 \eta^2 R^{-2.1}$ in the region of interest (i.e. $1\lesssim R/R_\mathrm{ms}\lesssim2.5$), and hence we have
\begin{equation}
P_\mathrm{rad}/P\simeq 5.6\times 10^3 \eta^2\alpha^{-1/2}m^{-5/2}\dot{m}^{2}R^{-5.9}.
\label{eq:pratio}
\end{equation}

\noindent Taking $P_\mathrm{rad}/P\sim 0.2$ leads to the radius of the outer edge of the unstable region, i.e.,

\begin{equation}
R_\mathrm{us}\simeq 4.3\eta^{0.34} \alpha^{-0.08} m^{-0.42}\dot{m}^{0.34}.
\label{eq:rus}
\end{equation}

For the unstable region, the $\dot{m}-\Sigma$ profile presents a S-shaped curve (where $\Sigma$ is the surface density of the disk), and the flow oscillates repeatly between the ``high state" (high $\dot{m}_{\mathrm{cr, h}}<\dot{m}<\dot{m}_{\mathrm{high}}$, radiation pressure and advective cooling dominate) and the ``low state" (low $\dot{m}_{\mathrm{low}}<\dot{m}<\dot{m}_{\mathrm{cr, l}}$, gas pressure and neutrino cooling dominate), corresponding to the upper branche and the lower branche in Figure \ref{fig:S-curve}, through a series of limt cycles one after another.  Note that the ``high state", ``unstable state" and ``low state" have the very solutions of region V, region IV and region III respectively (refer to \cite{XLW16} for the details), the expressions of the surface density of the disk for these branches are listed as follow:

\begin{subeqnarray}
\Sigma^{\mathrm{HS}}=1.3\times10^{17}A^{-2}B^{3/2}C^{1/2} \alpha^{-1}m^{-1} \ \ \ \ \ \ \ \ \ \ \ \nonumber \\
\times \dot{m} R_\mathrm{us}^{-1/2}\mathrm{\ \ \ g\ cm^{-2}},\\
\Sigma^{\mathrm{US}}=1.0\times10^{15}A^{-2/3}B^{7/6}C^{1/2}\mathscr{D}^{-1} \alpha^{-1/3}m^{5/3} \nonumber\\
\times \dot{m}^{-1}R_\mathrm{us}^{5/2} \mathrm{\ \ \ g\ cm^{-2}},\\
\Sigma^{\mathrm{LS}}=2.4\times10^{17}A^{-4/3}B^{2/3}C^{1/3}\mathscr{D}^{1/3} \alpha^{-2/3} \ \ \ \ \ \ \nonumber\\
\times \dot{m}^{1/3} \mathrm{\ \ \ g\ cm^{-2}}.
\label{eq:sigmas}
\end{subeqnarray}

In each cycle, the accretion rate will decrease from the high accretion rate $\dot{m}_{\mathrm{high}}$ to the higher critical value $\dot{m}_{\mathrm{cr, h}}$, and suddenly drop to the low accretion rate $\dot{m}_{\mathrm{low}}$. The flow then enters the ``low state", its accretion rate will gradually increase due to the mass feeding from the upper-stream (i.e., region III), until reaching the lower critical value $\dot{m}_{\mathrm{cr, l}}$. It jumps back directly to ``high state'' with accretion rate $\dot{m}_{\mathrm{high}}$.
Modulated by a series of limit cycles, the luminosity of the jet varies, giving the temporal behaviour of GRB prompt emission. From equation (\ref{eq:sigmas}), we get the four characteristic values of accretion rate in the S-shaped curve, i.e.,
\begin{subeqnarray}
&&\dot{m}_{\mathrm{high}}=4.7\times10^{-1} A^{5/6}B^{-17/24}C^{-1/8} \alpha^{5/12}m^{17/12}R_\mathrm{us}^{9/8}\nonumber \\
&&\ \ \ \ \ \ \ \ \sim 5.8\times 10^{-1}\alpha^{5/12}m^{17/12}R_\mathrm{us}^{1.0}\\
&&\dot{m}_{\mathrm{cr, h}}=8.8\times10^{-2} A^{2/3}B^{-1/6}\mathscr{D}^{-1/2} \alpha^{1/3}m^{4/3}R_\mathrm{us}^{3/2} \nonumber\\
&&\ \ \ \ \ \ \ \sim 7.6\times 10^{-2} \eta^{-0.5} \alpha^{1/3}m^{4/3}R_\mathrm{us}^{1.8} \\
&&\dot{m}_{\mathrm{cr, l}}=1.6\times10^{-2} A^{1/2}B^{3/8}C^{1/8}\mathscr{D}^{-1} \alpha^{1/4}m^{5/4}R_\mathrm{us}^{15/8} \nonumber\\
&&\ \ \ \ \ \ \ \sim 1.1\times10^{-2}\eta^{-1}\alpha^{1/4}m^{5/4}R_\mathrm{us}^{2.6}\\
&&\dot{m}_{\mathrm{low}}=1.1\times10^{-4} B^{2}C^{1/2}\mathscr{D}^{-5/2} m R_\mathrm{us}^{3} \nonumber \\
&&\ \ \ \ \ \ \ \sim 5.5\times 10^{-5}\eta^{-2.5}m R_\mathrm{us}^{4.6}
\label{eq:dotm}
\end{subeqnarray}

Now, we evaluate the variability timescale due to each limit cycle, which is taken as the minimum temporal variability ($\rm MTS$). The timescale of ``unstable state" can be ignored, and state transition is assumed to take place immediately once the accretion rate reaches the critical values. The duration of each limit cycle mainly depends on the evolution of the mass accretion rate, namely the viscous timescale in the ``low state" and the ``high state",
\begin{subequations}
\begin{eqnarray}
&& t_\mathrm{vis}^{\text{HS}}= 8.9\times10^{-6}C\mathscr{D}^{-1}\alpha^{-1}mR^{3/2} \mathrm{\ \ \ s},\\
&& t_\mathrm{vis}^{\text{LS}}= 1.7\times10^{-5}A^{2/3}B^{-5/6}C^{5/6}\mathscr{D}^{-2/3}\alpha^{-2/3}m^2 \nonumber\\
&& \ \ \ \ \  \times\ \dot{m}_{\mathrm{low}}^{-2/3}R^2 \mathrm{\ \ \ s},
\end{eqnarray}
\label{eq:t_vis}
\end{subequations}
\noindent where the viscous timescale is estimated by $t_\mathrm{vis}\sim r^2/\nu=(\alpha\Omega_\text{k})^{-1}(h/r)^{-2}$, here $\Omega_\text{k}\equiv\sqrt{GM/r^3}$.

Generally we have $t_\mathrm{vis}^{\text{LS}} > t_\mathrm{vis}^{\text{HS}}$, i.e., MTS is dominated by the timescale in ``low state" of the limit cycle $\mathrm{MTS} \sim t_\mathrm{vis}^{\text{LS}} \propto  \dot{m}_\mathrm{low}^{-2/3}$.

Substituting equation (\ref{eq:rus}) into equation (\ref{eq:dotm}d), we have the relation between the low state accretion rate $\dot{m}_{\mathrm{low}}$ of the unstable region and the disk accretion rate $\dot{m}$ as
\begin{equation}
\dot{m}_{\mathrm{low}}\sim 4.5\times10^{-2} \eta^{-0.94}\alpha^{-0.37} m^{-0.93} \dot{m}^{1.56}.
\label{eq:mdot_low}
\end{equation}

\noindent Combining equation (\ref{eq:t_vis}b) and (\ref{eq:mdot_low}) , we approximately have

\begin{eqnarray}
&&\mathrm{MTS}\sim 1.3\times10^{-4} A^{2/3}B^{-5/6}C^{5/6}\mathscr{D}^{-2/3}\eta^{0.63}\alpha^{-0.42}\nonumber \\
&&\ \ \ \ \ \ \ \times\ m^{2.62}\dot{m}^{-1.04}R^2 \mathrm{\ \ \ s},
\label{eq:MTS}
\end{eqnarray}
$t_{\rm vis}$ under the typical parameters is about $100\ \text{ms}$, which is in the same order of magnitude of the observations (e.g., \citealt{Mac13,GB14}).

\subsection{The jet driven by neutrino annihilation}
The neutrino annihilation ($\nu \bar{\nu} \rightarrow e^{+} e^{-}$) process above an NDAF can launch a relativistic jet reaching the GRB luminosity. An approximate expression for the neutrino annihilation power $\dot{E}_{\nu\bar{\nu}}$ is given by Zalamea \& Beloborodov (2011) as,
\begin{equation}
\dot{E}_{\nu \bar{\nu}} \simeq 6.2 \times 10^{49} \left(\frac{R_{\rm ms} }{2} \right)^{-4.8} \left(\frac{m}{3}\right)^{-3/2} \dot{m}_{-1}^{9/4} ~{\rm erg \ s^{-1}}.
\label{eq:Evv}
\end{equation}

Neutrino heating via neutrino absorption on baryons ($p+\bar{\nu}_e \rightarrow n+e^+$ and $n+\nu_e \rightarrow p+e^-$) in the atmosphere of NDAF can drive a baryonic wind (e.g., \citealt{MPQ08}). Since the majority of the mass lies in large radii, the main part of the wind originates from the region that is dominated by gas pressure and URCA cooling (Region II). According to \cite{Lei13}, the neutrino-heating driven baryon loading rate of the jet can be estimated as

\begin{eqnarray}
\dot{M}_{\rm j, \nu\bar{\nu}}
& = & 7.0 \times 10^{-7} A^{1.13} B^{-1.35} C^{0.22} \theta_{\rm j,-1}^2  \alpha_{-1}^{0.57} \epsilon_{-1}^{1.7} \nonumber \\
& \times & \left( \frac{R_{\rm ms} }{2} \right) ^{0.32} \dot{m}_{-1}^{1.7} \left(\frac{m}{3}\right)^{-0.9} \left(\frac{\xi}{2} \right)^{0.32} M_{\sun} {\rm s}^{-1},
\end{eqnarray}

\noindent where $\xi \equiv r/r_{\rm ms}$ is the disk radius in terms of $r_{\rm ms}$, $\epsilon\simeq(1-E_\mathrm{ms})$ denotes the neutrino emission efficiency, $E_\mathrm{ms}=(4\sqrt{R_\mathrm{ms}}-3a_\bullet)/\sqrt{3}R_\mathrm{ms}$ is the specific energy at ISCO, and $\theta_{\rm j}$ is the jet opening angle.

If most neutrino annihilation energy is converted into the kinetic energy of baryons after acceleration, the ultimate Lorentz factor of the jet is determined by the dimensionless ``entropy'' parameter, $\eta_0$, i.e.,
\begin{eqnarray}
\Gamma_{\rm max}  & \simeq & \eta_0 \equiv \dot{E}_{\nu \bar{\nu}}/\dot{M}_{\rm j, \nu\bar{\nu}} c^2 \nonumber \\
& =  & 50 A^{-1.13} B^{1.35} C^{-0.22} \theta_{\rm j,-1}^{-2} \alpha_{-1}^{-0.57} \epsilon_{-1}^{-1.7} \left(\frac{\xi}{2}\right)^{-0.32} \nonumber \\
& \times & \left(\frac{ R_{ms}}{2}\right)^{-5.12} \left(\frac{m}{3}\right)^{-0.6} \dot{m}_{-1}^{0.58}
\label{eq:eta}.
\end{eqnarray}

Inspecting Eqs. (\ref{eq:MTS}), (\ref{eq:Evv}) and (\ref{eq:eta}), one finds MTS,  $\dot{E}_{\nu \bar{\nu}}$ and $\eta_0$ are functions of $\dot{m}$, $m$ and $a_*$. Usually the BH mass in GRBs $m$ varies in a narrow range of (2.5, 10) (e.g. \citealt{PWF99}). Considering the hyper-accreting process during the prompt emission phase, the BH is quickly spun up, so that the $a_*$-dependence is not significant (\citealt{Lei13}). Therefore, the $\dot{m}$-dependence may be the key to define the ${\rm MTS} - \eta_0$ correlation (${\rm MTS} \propto \eta_0^{-1.8} $) and ${\rm MTS} - \dot{E}_{\nu \bar{\nu}}$ correlation (${\rm MTS} \propto \dot{E}_{\nu \bar{\nu}}^{-0.5}$). Comparing with the data, the predicted indices 1.8 (0.5) in ${\rm MTS} - \eta_0$ correlation (${\rm MTS} - \dot{E}_{\nu \bar{\nu}}$ correlation) are significantly smaller than the observational value $4.8\pm1.5$ ($1.0\pm0.1$).

\begin{figure*}[h!t!]
\centering
\includegraphics[width=85mm]{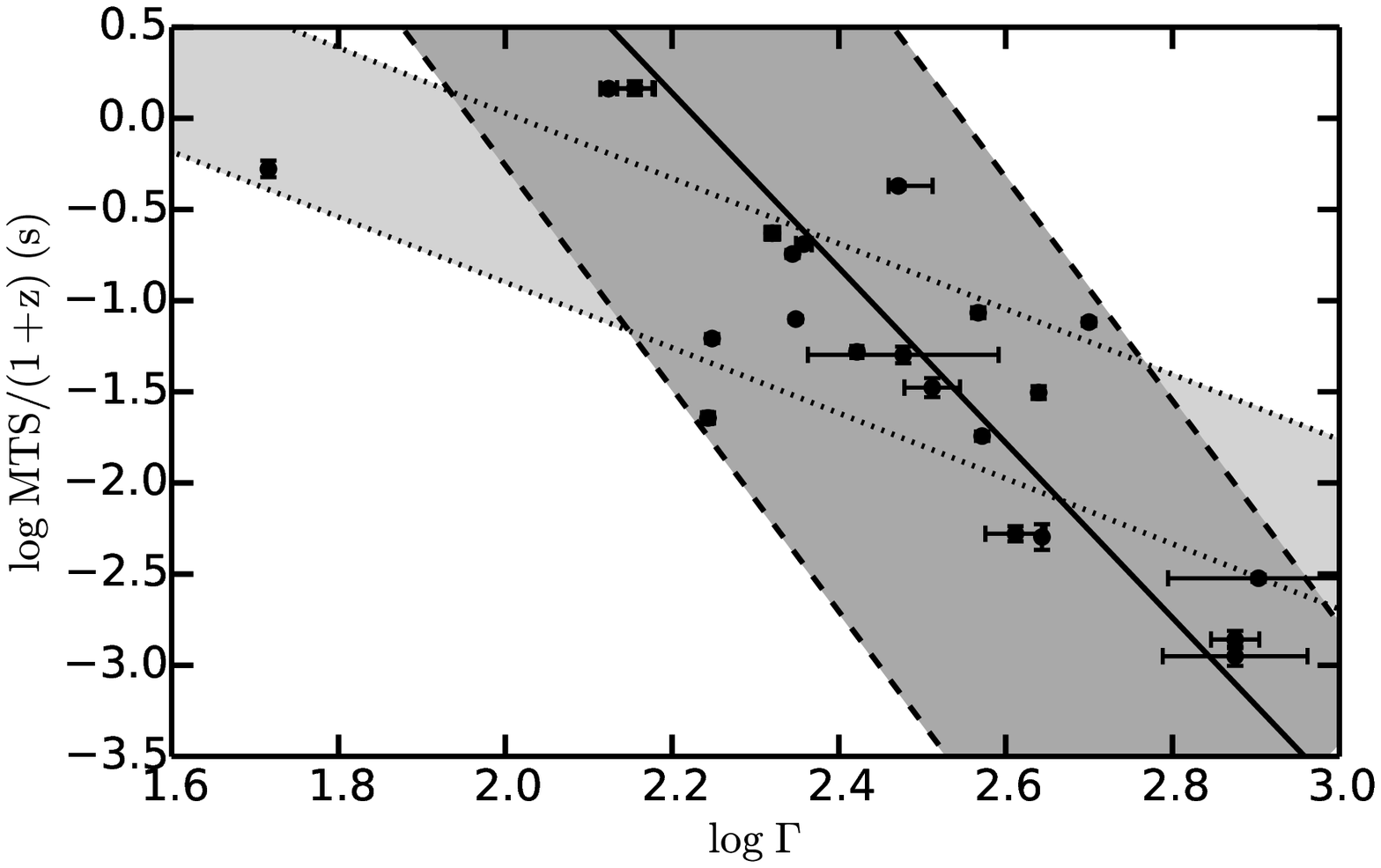}
\includegraphics[width=85mm]{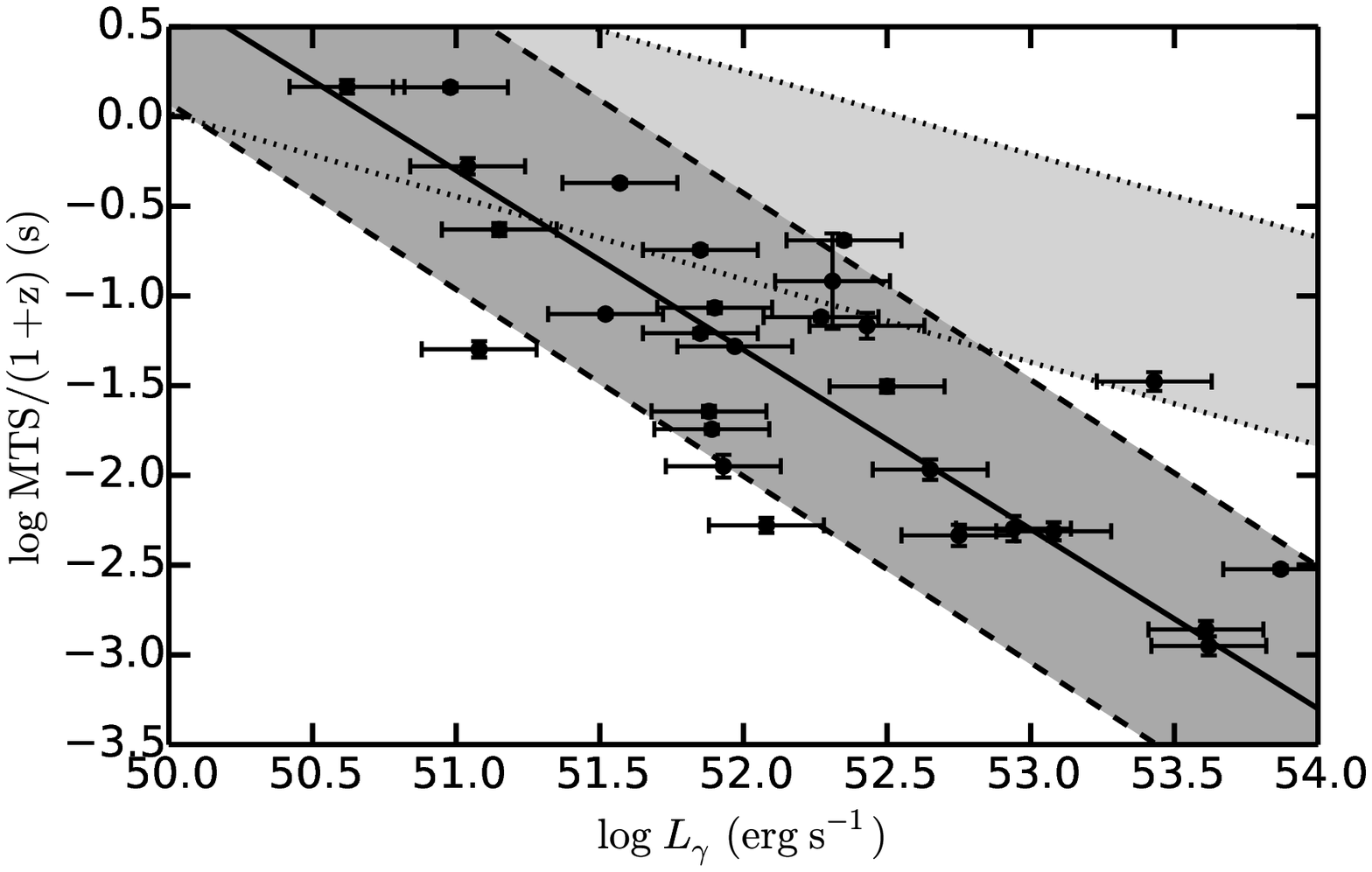}
\caption{Left panel: ${\rm MTS}$ vs. $\Gamma$; Right panel: ${\rm MTS}$ vs. $L_{\gamma}$. The black dots are GRB data adopted from \cite{Wu16a}. The solid lines are best fits to the data: ${\rm MTS/(1+z)} \propto \Gamma^{-4.8} $ and ${\rm MTS/(1+z)} \propto L_{\gamma}^{-1.0} $. The predictions from the $\nu \bar\nu$ - annihilation mechanism (light shaded regions between the dotted lines) and the BZ mechanism (shaded regions between the dashed lines) are shown for comparison, in which $\dot{m}$ changes in a wide range with the other parameters fixed. For $\nu \bar\nu$ - annihilation mechanism, we take $a_\bullet=0.8$, $\alpha=0.1$, $\eta=1$, $R=2.5$ denoting region IV and 20 denoting region II, $\theta_{\rm j}=0.03$, $\eta_\gamma=0.01$, here $\eta_\gamma\equiv L_\gamma(1-\cos\theta_{\rm j})/\dot{E}_{\nu \bar{\nu}}$. The top (bottom) dotted lines correspond to the cases with $m=10$ (2.5). While for BZ mechanism, we take $a_\bullet=0.2$, $\alpha=0.2$, $\eta=1$, $R=6$ denoting region IV and 20 denoting region II, $r_z=5\times10^{11}\mathrm{\ cm}$, $f_\mathrm{p}=0.01$, $\theta_{\rm j}=0.4$, $\theta_{\rm B}=0.011$, $\eta_\gamma=0.01$, here $\eta_\gamma\equiv L_\gamma(1-\cos\theta_{\rm j})/\dot{E}_{\rm BZ}$. The top (bottom) dashed lines correspond to  the cases with $m=10$ (2.5).}
\label{fig:data}
\end{figure*}

However, the m-dependence will add to the scatter in the correlations. In Fig. \ref{fig:data}, we plot MTS versus $\Gamma$ (left), and MTS versus $L_{\gamma}$ (right) with dotted lines for $m=2.5$ (bottom) and 10 (top). For each $m$ case, we change ${\dot m}$ in a wide range of values and keep other parameters to fixed values, see Fig. \ref{fig:data}. If we allow $m$ to randomly vary in the range of (2.5, 10), the simulated GRBs should be scattered in the light shaded region between the dotted lines. From Fig. \ref{fig:data}, we find that the predictions with NDAF model are inconsistent with the data.

\subsection{The jet driven by BZ mechanism}
We now consider BZ scenario for the jet production. The magnetic field in BZ mechanism is established and supported by the magnetized NDAF around the BH, and the baryons in the jet are also loaded from the neutrino-driven wind. Unlike the NDAF scenario, the existence of the magnetic barrier will significantly suppress the baryon loading from the disk (e.g. \citealt{Li00}). Therefore, a baryon-poor jet will be built in BZ model.

The BZ power from a BH with mass $M_{\bullet}$ and spin $a_\bullet$ is (e.g. \citealt{Lee00, Lei13, Wu13})
\begin{equation}
\dot{E}_{\rm BZ}=9.3 \times 10^{52} a_\bullet^2 \dot{m}_{-1}  X(a_\bullet) \ {\rm erg \ s^{-1}} ,
\label{eq:Lmag}
\end{equation}
where $X(a_\bullet)=F(a_\bullet)/(1+\sqrt{1-a_\bullet^2})^2 $, $ F(a_{\bullet})=[(1+q^2)/q^2][(q+1/q) \arctan q-1]$, and $q= a_{\bullet} /(1+\sqrt{1-a^2_{\bullet}})$.

As for the baryon loading process, with considering the blocking effect of the magnetic barrier on the protons, \cite{Lei13} suggested the neutron drift rate into the jet as

\begin{eqnarray}
\dot{M}_\text{j,BZ}&\simeq& 3.5\times10^{-7}A^{23/30}B^{-33/40}C^{7/120}f_\text{p,-1}^{-1/2}\theta_\text{j,-1}
\theta_\text{B,-2}^{-1} \nonumber\\
&\times& \alpha_{-1}^{23/60}\epsilon_{-1}^{5/6} \dot{m}_{-1}^{5/6}\left(\frac{R_\mathrm{ms}}{2}\right)^{1/120}\left(\frac{m}{3}\right)^{-11/20}r_{z,11}^{1/2} \nonumber\\
&\times& \left(\frac{\xi}{2}\right)^{1/120} \ M_\odot\mathrm{\ s^{-1}}.
\label{eq:Mjbz2}
\end{eqnarray}
\noindent where $f_\mathrm{p}$ denotes the fraction of the protons in the wind, $r_z$ is the distance from the BH in the jet direction, $\theta_\mathrm{B}$ is introduced here to reflect the fact that only the protons with small ejected angle ($\lesssim \theta_\mathrm{B}$) with respect to the field lines can come into the disk atmosphere.

For such a magnetized central engine, the maximum available energy per baryon in the jet can be evaluated as

\begin{eqnarray}
\mu_0&\simeq&\frac{\dot{E}_\text{BZ}}{\dot{M}_\text{j,BZ}c^2}=1.5\times10^5 A^{-23/30}B^{33/40}C^{-7/120} f_\text{p,-1}^{1/2}\nonumber\\
&\times& \theta_\text{j,-1}^{-1} \theta_\text{B,-2} \alpha_{-1}^{-23/60}\epsilon_{-1}^{-5/6}r_{z,11}^{-1/2} a_\bullet^2 X(a_\bullet) \nonumber\\
&\times& \left(\frac{R_\mathrm{ms}}{2}\right)^{-1/120}\left(\frac{\xi}{2}\right)^{-1/120}\left(\frac{m}{3}\right)^{11/20} \dot{m}_{-1}^{1/6}.
\label{eq:mu0}
\end{eqnarray}

Since the acceleration process of the jet suffers from uncertainties, the terminating Lorentz factor of the jet generally satisfies
\begin{equation}
 \Gamma_{\rm min} < \Gamma < \Gamma_{\rm max},
\label{gamma}
\end{equation}
with the specific value depending on the detailed process of magnetic dissipation, such as the magnetic energy relaxion through the Internal-Collision induced MAgnetic Reconnection and Turbulence (ICMART) \citep{ZY11} or the shearing interaction at the inner/outer layer interface in the possible two-componet jet (e.g. \citealt{Wang14}). Following \cite{Lei13}, we separately take $\Gamma_{\rm min}=\max(\mu_{0}^{1/3},\eta_0)$ ($\eta_0=\dot{E}_{\nu \bar{\nu}} /(\dot{M}_{\rm j, BZ} c^2)$) and $\Gamma_{\rm max} =  \mu_0$, corresponding to the start and the end of the slow acceleration phase in a hybrid outflow (a detailed discussion of the acceleration dynamics of an arbitrarily magnetized relativistic or hybrid jet is referred to \cite{GZ15}).

Based on Eqs. (\ref{eq:MTS}), (\ref{eq:Lmag}) and (\ref{eq:mu0}), we find the  ${\rm MTS} - \mu_0$ correlation (${\rm MTS} \propto \mu_0^{-6.2} $) and ${\rm MTS} - \dot{E}_{\rm BZ}$ correlation (${\rm MTS} \propto \dot{E}_{\rm BZ}^{-1.04}$). With the consideration of the spin-up process due to accretion and the spin-down process due to the BZ process, the BH spin parameter always evolves to an equilibrium value (\citealt{Lei05}), so that the $a_*$-dependence essentially does not enter the problem. As shown in Fig. \ref{fig:data}, the $m$-dependence will add to the scatter in the correlations. The top (bottom) dashed lines represent the MTS-$\Gamma$ (left) and MTS-$L_{\gamma}$ correlations (right) for $m=10$ (2.5). GRBs with BH mass in the range of (2.5, 10) should be scattered in the shaded region between the dashed lines. We can find that the predictions from the BZ mechanism are more consistent with the empirical correlations than that of $\nu \bar\nu$ - annihilation mechanism.


\section{Conclusions and Discussions}
In this paper, we compare the theoretical predictions from neutrino annihilation and BZ processes with the empirical  ${\rm MTS} - \Gamma$ and ${\rm MTS} - L_{\gamma}$ correlations. We find that both empirical correlations favor the BZ scenario. These correlations may bring us a clue to the GRB central engine, i.e., a good fraction of GRBs may be driven by a hyperaccretion system consisting of a stellar BH and a surrounding NDAF. Furthermore, the jet power of GRBs may be supplied by the BH rotating energy through the BZ mechanism. This result is consistent with the implication from the $L_\gamma-E_\mathrm{p,z}-\Gamma_0$ correlaton proposed by \cite{Liang15}, which suggested that the GRB jet might be Poynting-flux-dominated. In addition, our result is also in agreement with \cite{Yi16}, which investigated the BH central engines from the correlation between $\Gamma$ and the beaming corrected luminosity. Even in BZ model, the disk is still NDAF. The latter plays two important roles in BZ scenario. Firstly, The hyper-accretion is necessary to maintain the strong magnetic field for BZ process. Secondly, the neutrino-heating wind in the surface of NDAF acts as a significant contribution to the jet baryon-loading.

In this work, we adopt $\dot{m}$ as the primary correlating variable. As mentioned in Section 2, the dependence of BH spin and mass may also enter the correlations. However, the BH spin parameter will quickly evolve to the maximum value (for NDAF model) or equilibrium value (for BZ scenario), so that the $a_*$ dependence is not important. On the other hand, the empirical correlations were obtained from the average $\Gamma$ and $L_\gamma$. In order to compare with the data one needs to calculate the average $\Gamma$, $\dot{E}_{\nu \bar{\nu}}$ and $\dot{E}_{\rm BZ}$, which smears the $a_*$ dependence (\citealt{Lei13}). The BH mass $m$ is generally believed to vary from 2.5 to 10 in GRBs (e.g., \citealt {PWF99}). Therefore, with $m$-dependence only, the theoretical models can hardly reproduce the wide range of the observed $L_\gamma$ and MTS. As shown in Fig. \ref{fig:data}, it adds to the scatter in the correlations, but does not change our main conclusion. A detailed study of the effects due to the dependences on the model parameters (e.g., $m$, $a_*$, $\alpha$, $\theta_{\rm j}$, $\eta_\gamma$) will be addressed in future work.

It is interesting that the BZ mechanism and NDAF process have mutual influence on each other. On one hand, the magnetic field in BZ mechanism is supported (or even established, e.g. \cite{Cao14}) by NDAF. On the other hand, such a magnetic filed has strong effects in suppressing the baron-loading from the neutrino-driven wind. The further discussion for such a co-existence relation is beyond this work.

In our model, we define $\rm MTS$ as the viscous timescale due to viscous instability. There are other possible disk origins for the temporal variation. \cite{Cao14} investigated the capability of the NDAF to drag the large scaled magnetic field inward the BH, the result shows that the competition between the inherent diffusion and the accretion driven advection of the magnetic field leads to a oscillating accretion as well as an episodic jet, the oscillation timescale can be about one second which is comparable with the observed variation in the soft extended emission of short GRBs. An alternative mechanism involving the temporal variation is referred to the disk's inertial-acoustic oscillation which has been extensively discussed (e.g. \citealt{Kato78, Wag01, YL15}). For aperiodic variation, some authors thought it reflects the possible propagating fluctuations in the disk (e.g., \citealt{Lyu97, Lin16}). It's worthwhile to check the feasibility of these different models in interpreting the $\rm MTS$ related correlations in the future.

\acknowledgments
We thank the anonymous referee for constructive comments and suggestions. We also thank Qing-wen Wu, Bing Zhang and Yuan-Chuan Zou for helpful discussion. The numerical calculations in this work were performed by using a high performance computing cluster (Hyperion) at HUST. This work is supported by the National Basic Research Program (`973' Program) of China (grants 2014CB845800), the National Natural Science Foundation of China under grants U1431124, 11361140349 (China-Israel jointed program).

\label{reference}

\end{document}